\documentclass[twocolumn,10pt]{IEEEtran}

\usepackage{amsmath,amssymb,amsthm}
\usepackage{graphicx}
\usepackage{xcolor}
\usepackage{subfigure}
\usepackage[acronym]{glossaries}

\newacronym{awgn}{AWGN}{additive white Gaussian noise}
\newacronym{bc}{BC}{broadcast channel}
\newacronym{em}{EM}{electromagnetic}
\newacronym{los}{LoS}{line-of-sight}
\newacronym{mimo}{MIMO}{multiple-input multiple-output}
\newacronym{miso}{MISO}{multiple-input single-output}
\newacronym{mac}{MAC}{multiple access channel}
\newacronym{nlos}{NLoS}{non-line-of-sight}
\newacronym{ris}{RIS}{reconfigurable intelligent surface}
\newacronym{simo}{SIMO}{single-input multiple-output}
\newacronym{snr}{SNR}{signal-to-noise ratio}
\newacronym{sre}{SRE}{smart radio environment}
\newacronym{ttd}{TTD}{true time delay}
\newacronym{ula}{ULA}{uniform linear array}

\begin{document}
\bstctlcite{BSTcontrol}

\title{\LARGE Capacity of Two-User Wireless Systems Aided by Movable Signals}

\author{Matteo~Nerini,~\IEEEmembership{Senior~Member,~IEEE},
        Bruno~Clerckx,~\IEEEmembership{Fellow,~IEEE}

\thanks{This work has been supported in part by UKRI under Grant EP/Y004086/1, EP/X040569/1, EP/Y037197/1, EP/X04047X/1, EP/Y037243/1.}
\thanks{Matteo Nerini and Bruno Clerckx are with the Department of Electrical and Electronic Engineering, Imperial College London, SW7 2AZ London, U.K. (e-mail: m.nerini20@imperial.ac.uk; b.clerckx@imperial.ac.uk).}}

\maketitle

\begin{abstract}
Movable signals have emerged as a third approach to enable \glspl{sre}, complementing \glspl{ris} and flexible antennas.
This paper investigates their potential to enhance multi-user wireless systems.
Focusing on two-user systems in \gls{los}, we characterize the capacity regions of the \gls{mac} and \gls{bc}.
Interestingly, movable signals can dynamically adjust the operating frequency to orthogonalize the user channels, thereby significantly expanding the capacity regions.
We also study frequency optimization, constraining it in a limited frequency range, and show that movable signals provide up to 45\% sum rate gain.
\end{abstract}

\glsresetall


\section{Introduction}

\Glspl{sre} promise major advances in wireless communications by allowing the channel to be shaped and treated as an optimization variable \cite{dir19,dir20}.
The main enabling technology of \glspl{sre} is \gls{ris}, extensively investigated in recent years \cite{wu21}.
A \gls{ris} consists of an array of reflecting elements with reconfigurable \gls{em} properties that can be tuned to manipulate impinging wavefronts.
By deploying a \gls{ris} in the environment and adjusting its \gls{em} properties, we can shape the propagation conditions, enhancing desired signals while mitigating interference.
\glspl{sre} enabled by \gls{ris} rely on modifying the \gls{em} characteristics of the environment and can therefore be referred to as \gls{em}-domain \glspl{sre}.

A second approach for enabling \glspl{sre} operates in the space domain, by adjusting the distances between transmit antennas, receive antennas, and nearby objects through flexible antennas.
Several flexible antenna technologies have been proposed, including fluid and movable antennas, which exploit the fact that even sub-wavelength displacements can lead to substantially different fading realizations in rich scattering environments \cite{won21,zhu24}.
To provide greater adaptability, pinching antennas allow the antennas to be repositioned along a waveguide \cite{din25}.

A third approach for enabling \glspl{sre} lies in the frequency domain.
Beyond the \gls{em} properties of objects in the environment and the spatial distances, the signal frequency also constitutes a controllable parameter of the wireless channel.
Recent work has introduced movable signals, a technique in which the signal spectrum can be ``moved'' along the frequency axis to reshape the channel \cite{ner25-1,ner25-2}.
Unlike conventional systems, where the frequency is selected within a relatively narrow range, with movable signals the frequency is reconfigured within a much wider range (e.g., $[f_{\text{min}},f_{\text{max}}]$ with $f_{\text{max}}=1.8f_{\text{min}}$), resulting in significant channel shaping capabilities.
Compared to \glspl{ris}, movable signals do not rely on reconfiguring the \gls{em} properties of surrounding surfaces but instead exploit the frequency selectivity of the environment.
Compared to flexible antennas, movable signals do not alter distances but instead the wavelength, inducing similar channel variations.
Movable signals therefore offer a third appealing approach for enabling \glspl{sre}, requiring the sub-bands available for transmission to be suitably distributed along the frequency spectrum (such as through carrier aggregation \cite{she12}), without the need for reconfigurable or movable hardware.

Previous work on movable signals has shown that dynamically reconfiguring the operating frequency enables effective beam-steering capabilities for serving a receiver in \gls{los} \cite{ner25-1,ner25-2}.
Inspiring works also studied the frequency-dependent beamforming capabilities of specific array architectures, such as leaky-wave antennas and arrays with \gls{ttd} devices \cite{gab23,zhu25,zha21,li22}.
This property has been leveraged to serve users by assigning each of them a distinct frequency, either via leaky-wave antennas \cite{gab23,zhu25} or via arrays including \gls{ttd} devices \cite{zha21,li22}.
However, serving multiple users via orthogonal frequency channels is known to be suboptimal.
How to simultaneously serve multiple users on the same frequency, assumed to be reconfigurable with high flexibility, remains less explored.

In this paper, we fill this gap by studying the fundamental limits of two-user systems aided by movable signals, in which the users operate on the same frequency.
\textit{First}, we characterize the capacity region of wireless systems where a multi-antenna base station serves two single-antenna users in \gls{los} using movable signals.
We show that movable signals can make the two channels orthogonal by suitably choosing the operating frequency, therefore achieving a highly enhanced capacity region.
\textit{Second}, we provide transmission strategies allowing us to exactly achieve the boundary of the capacity region, considering both the uplink and the downlink, namely the \gls{mac} and the \gls{bc}.
\textit{Third}, we show how to reconfigure the operating frequency when it is constrained within a limited frequency range.
Numerical results show that, even under a constrained frequency range, movable signals can approach the sum-rate capacity and significantly outperform conventional systems.
In contrast to conventional frequency allocation over a narrow range that cannot meaningfully alter the \gls{los} channel, movable signals enable channel orthogonalization and substantial capacity gains.

\section{System Model}
\label{sec:model}

In this section, we present the system model.
We first introduce the two-user uplink and downlink transmission models, also referred to as \gls{mac} and \gls{bc}, respectively, and then describe the wireless channel model, assumed to be \gls{los} as it is relevant in millimeter waves and terahertz communications.

\begin{figure}[t]
\centering
\includegraphics[width=0.22\textwidth]{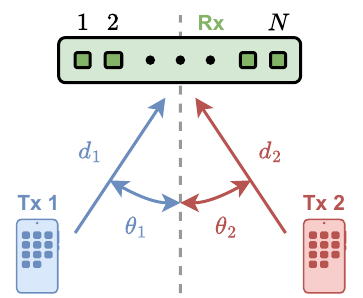}
\caption{Two-user multiple access channel (MAC).}
\label{fig:system}
\end{figure}


For the uplink, we consider a \gls{mac} where $K=2$ single-antenna users transmit to an $N$-antenna base station, as illustrated in Fig.~\ref{fig:system}.
The transmitted signals are denoted as $x_{ul,k}\in\mathbb{C}$, and are subject to the power constraints $\mathbb{E}[\vert x_{ul,k}\vert^2]=P_k$, where $P_k$ is the transmitted signal power of user $k$, for $k=1,2$.
Therefore, the received signal $\mathbf{y}_{ul}\in\mathbb{C}^{N\times1}$ is given by
\begin{equation}
\mathbf{y}_{ul}=\sum_{k=1}^2\mathbf{h}_{ul,k}x_{ul,k}+\mathbf{n}_{ul},
\end{equation}
where $\mathbf{h}_{ul,k}\in\mathbb{C}^{N\times 1}$ is the wireless channel between user $k$ and the base station, for $k=1,2$, and $\mathbf{n}_{ul}\in\mathbb{C}^{N\times 1}$ is the \gls{awgn} at the base station with noise power $\sigma^2$.

For the downlink, we consider the corresponding \gls{bc}, where an $N$-antenna base station transmits to $K=2$ single-antenna users.
The transmitted signal is denoted as $\mathbf{x}_{dl}\in\mathbb{C}^{N\times 1}$, and is subject to the power constraint $\mathbb{E}[\Vert\mathbf{x}_{dl}\Vert^2]=P$, where $P$ is the transmitted signal power.
In general, we have $\mathbf{x}_{dl}=\mathbf{x}_{dl,1}+\mathbf{x}_{dl,2}$, where $\mathbf{x}_{dl,1},\mathbf{x}_{dl,2}\in\mathbb{C}^{N\times 1}$ are two statistically independent signals intended for the two users, subject to $\mathbb{E}[\Vert\mathbf{x}_{dl,1}\Vert^2]=P_1$ and $\mathbb{E}[\Vert\mathbf{x}_{dl,2}\Vert^2]=P_2$, with $P_1+P_2=P$.
Accordingly, the received signal at user $k$, which is denoted as $y_{dl,k}\in\mathbb{C}$, writes as
\begin{equation}
y_{dl,k}=\mathbf{h}_{dl,k}\mathbf{x}_{dl}+n_{dl,k},
\end{equation}
where $\mathbf{h}_{dl,k}\in\mathbb{C}^{1\times N}$ is the wireless channel between the base station and user $k$, and $n_{dl,k}\in\mathbb{C}$ is the \gls{awgn} at user $k$ with noise power $\sigma^2_k$, for $k=1,2$.


To model the wireless channels $\mathbf{h}_{ul,k}$ and $\mathbf{h}_{dl,k}$, we assume that the antenna array at the base station is a \gls{ula} located along the $x$-axis and centered at $x=0$, such that the $n$th transmitting antenna has $x$ coordinate
$x_n=(n-(N+1)/2)d_A$,
where $d_A$ is the antenna spacing.
We denote as $d_k$ the distance between user $k$ and the center of the base station array, and as $\theta_k\in[-\pi/2,\pi/2]$ the angle of the user $k$ direction with respect to the base station array normal, as shown in Fig.~\ref{fig:system}.
Hence, the distance between user $k$ and antenna $n$ at the base station is
\begin{equation}
d_{n,k}=d_k-\left(n-\frac{N+1}{2}\right)d_A\sin\left(\theta_k\right),\label{eq:dnk}
\end{equation}
assuming far-field propagation.
Depending on $d_{n,k}$, the entries of the uplink channels $\mathbf{h}_{ul,1}$ and $\mathbf{h}_{ul,2}$ are given by $\left[\mathbf{h}_{ul,k}\right]_n=e^{-j\frac{2\pi}{\lambda}d_{n,k}}$, where $\lambda=c/f$ is the wavelength, $c$ the speed of light, and $f$ the frequency.
Following \eqref{eq:dnk}, we can write more explicitly
\begin{equation}
\left[\mathbf{h}_{ul,k}\right]_n
=e^{-j\frac{2\pi}{\lambda}\left[d_k-\left(n-\frac{N+1}{2}\right)d_A\sin\left(\theta_k\right)\right]},\label{eq:h}
\end{equation}
for $n=1,\ldots,N$ and $k=1,2$.
In a system aided by movable signals, the frequency $f$ can be dynamically optimized to shape the wireless channels.
The downlink channels are $\mathbf{h}_{dl,k}=\mathbf{h}_{ul,k}^T$, for $k=1,2$, assuming that channel reciprocity holds between the uplink and the downlink.

\section{Capacity of Multiple Access Channel}
\label{sec:mac}

In this section, we derive the capacity region of the two-user \gls{mac} aided by movable signals described in Section~\ref{sec:model}.
This region is defined as the set of all rate pairs $(R_1,R_2)$ such that users $1$ and $2$ can simultaneously and reliably communicate at rates $R_1$ and $R_2$, respectively \cite[Chapter~12]{cle13}.

\subsection{Characterization of the Capacity Region}

We begin by noticing that the rate of user $k$ is upper bounded by its rate in a single-user \gls{simo} system, i.e.,
\begin{align}
R_k
&\leq\log_2\det\left(\mathbf{I}_{N}+\frac{P_k}{\sigma^2}\mathbf{h}_{ul,k}\mathbf{h}_{ul,k}^H\right)\\
&=\log_2\left(1+\frac{P_k}{\sigma^2}\mathbf{h}_{ul,k}^H\mathbf{h}_{ul,k}\right)\label{eq:MAC-Rk1}\\
&=\log_2\left(1+\frac{P_k}{\sigma^2}N\right),\label{eq:MAC-Rk}
\end{align}
for $k=1,2$, where in \eqref{eq:MAC-Rk1} we used Sylvester's determinant theorem and in \eqref{eq:MAC-Rk} we exploited $\mathbf{h}_{ul,k}^H\mathbf{h}_{ul,k}=\Vert\mathbf{h}_{ul,k}\Vert^2=N$.
In addition, the sum rate $R_1+R_2$ is upper bounded by the rate achievable when the two users ``cooperate'' forming a two-antenna array, where antenna $k$ has a power constraint $P_k$, for $k=1,2$.
By introducing the \gls{mimo} channel $\mathbf{H}_{ul}=[\mathbf{h}_{ul,1},\mathbf{h}_{ul,2}]\in\mathbb{C}^{N\times2}$ and the transmit covariance matrix $\mathbf{Q}_{ul}=\text{diag}(P_1,P_2)\in\mathbb{C}^{2\times2}$, such an upper bound is given by
\begin{align}
R_1+R_2
&\leq\log_2\det\left(\mathbf{I}_{N}+\frac{1}{\sigma^2}\mathbf{H}_{ul}\mathbf{Q}_{ul}\mathbf{H}_{ul}^H\right)\\
&=\log_2\det\left(\mathbf{I}_{2}+\frac{1}{\sigma^2}\mathbf{Q}_{ul}\mathbf{H}_{ul}^H\mathbf{H}_{ul}\right).\label{eq:MAC-sum}
\end{align}
Therefore, the capacity region is bounded by the constraints in \eqref{eq:MAC-Rk} and \eqref{eq:MAC-sum}.

To enhance the capacity region with movable signals, we observe that the upper bound in \eqref{eq:MAC-sum} is globally maximized when the channels $\mathbf{h}_{ul,1}$ and $\mathbf{h}_{ul,2}$ are orthogonal, i.e., when $\mathbf{h}_{ul,2}^H\mathbf{h}_{ul,1}=0$.
In this case, \eqref{eq:MAC-sum} can be rewritten as
\begin{align}
R_1+R_2
&\leq\log_2\det\left(\mathbf{I}_{2}+\frac{1}{\sigma^2}
\begin{bmatrix}
P_1N & 0\\
0 & P_2N
\end{bmatrix}
\right)\\
&=\log_2\left(1+\frac{P_1N}{\sigma^2}\right)+\log_2\left(1+\frac{P_2N}{\sigma^2}\right),
\end{align}
becoming equivalent to \eqref{eq:MAC-Rk}.
Therefore, if it were possible to reconfigure the signal frequency such that $\mathbf{h}_{ul,2}^H\mathbf{h}_{ul,1}=0$, the capacity region would be bounded solely by \eqref{eq:MAC-Rk}.

Interestingly, it is possible to orthogonalize the channels $\mathbf{h}_{ul,1}$ and $\mathbf{h}_{ul,2}$ by reconfiguring the signal frequency, as shown in the following.
Since we want to set the frequency such that $\mathbf{h}_{ul,2}^H\mathbf{h}_{ul,1}=0$, we begin by explicitly writing $\mathbf{h}_{ul,2}^H\mathbf{h}_{ul,1}$ as a function of the wavelength $\lambda$ as
\begin{multline}
\mathbf{h}_{ul,2}^H\mathbf{h}_{ul,1}
=\sum_{n=1}^N\left(e^{j\frac{2\pi}{\lambda}\left[d_2-\left(n-\frac{N+1}{2}\right)d_A\sin(\theta_2)\right]}\right.\\
\left.\times e^{-j\frac{2\pi}{\lambda}\left[d_1-\left(n-\frac{N+1}{2}\right)d_A\sin(\theta_1)\right]}\right),
\end{multline}
where we applied \eqref{eq:h}, which can be simplified as
\begin{equation}
\mathbf{h}_{ul,2}^H\mathbf{h}_{ul,1}
=e^{j\alpha}\sum_{n=1}^Ne^{j\frac{2\pi}{\lambda}nd_A\left(\sin\left(\theta_1\right)-\sin\left(\theta_2\right)\right)},\label{eq:hh1}
\end{equation}
where $\alpha$ is a constant independent on the index $n$, defined as
$\alpha=2\pi(d_2-d_1+(N+1)d_A(\sin(\theta_2)-\sin(\theta_1))/2)/\lambda$.
Looking at \eqref{eq:hh1}, we notice that since $e^{j\alpha}\neq0$ for any value of $\lambda$, we have $\mathbf{h}_{ul,2}^H\mathbf{h}_{ul,1}=0$ if and only if
\begin{equation}
\sum_{n=1}^Ne^{j\frac{2\pi}{\lambda}nd_A\left(\sin\left(\theta_1\right)-\sin\left(\theta_2\right)\right)}=0,
\end{equation}
or, equivalently,
\begin{equation}
\sum_{n=0}^{N-1}e^{j\frac{2\pi}{\lambda}nd_A\left(\sin\left(\theta_1\right)-\sin\left(\theta_2\right)\right)}=0.\label{eq:hh2}
\end{equation}
Therefore, we can exploit the property of finite geometric series $\sum_{n=0}^{N-1}r^n=(1-r^N)/(1-r)$, for any $r\neq1$, to rewrite \eqref{eq:hh2} as
\begin{equation}
\frac
{1-e^{j\frac{2\pi}{\lambda}Nd_A\left(\sin\left(\theta_1\right)-\sin\left(\theta_2\right)\right)}}
{1-e^{j\frac{2\pi}{\lambda}d_A\left(\sin\left(\theta_1\right)-\sin\left(\theta_2\right)\right)}}=0,\label{eq:hh3}
\end{equation}
which is satisfied under two conditions.

First, the numerator in \eqref{eq:hh3} must be zero, i.e., we must have
\begin{equation}
\frac{2\pi}{\lambda}Nd_A\left(\sin\left(\theta_1\right)-\sin\left(\theta_2\right)\right)=2\pi L,
\end{equation}
with $L\in\mathbb{Z}$.
The optimal wavelength should therefore be set as a function of the angles $\theta_1$ and $\theta_2$ as $\lambda^\star=Nd_A\vert\sin(\theta_1)-\sin(\theta_2)\vert/L$, with $L\in\mathbb{N}$.
Equivalently, the signal frequency should be set as
\begin{equation}
f^\star=\frac{Lf_A}{N\left\vert\sin\left(\theta_1\right)-\sin\left(\theta_2\right)\right\vert},\label{eq:f}
\end{equation}
with $L\in\mathbb{N}$, where we have introduced $f_A=c/d_A$, indicating that there are infinitely many possible values of $f^\star$, depending on the values of $L$.

Second, the denominator in \eqref{eq:hh3} must be different from zero, i.e., we need
\begin{equation}
\frac{2\pi}{\lambda}d_A\left(\sin\left(\theta_1\right)-\sin\left(\theta_2\right)\right)\neq2\pi M,
\end{equation}
with $M\in\mathbb{Z}$, which requires $\theta_2\neq\theta_1$ and
\begin{equation}
f^\star\neq\frac{Mf_A}{\left\vert\sin\left(\theta_1\right)-\sin\left(\theta_2\right)\right\vert},\label{eq:f-neq}
\end{equation}
with $M\in\mathbb{N}$.
This means that the two channels can be made orthogonal by adjusting the frequency only when $\theta_2\neq\theta_1$, which is intuitive since if the channels are perfectly aligned, they remain aligned for any frequency value.
Besides, the optimal frequency must be a multiple of $f_A/(N\vert\sin(\theta_1)-\sin(\theta_2)\vert)$ but not a multiple of $f_A/\vert\sin(\theta_1)-\sin(\theta_2)\vert$, otherwise the $N$ additive complex terms in $\mathbf{h}_{ul,2}^H\mathbf{h}_{ul,1}$ have all the same phase and $\vert\mathbf{h}_{ul,2}^H\mathbf{h}_{ul,1}\vert=N$.

\subsection{Achievability of the Capacity Region}

\begin{figure}[t]
\centering
\includegraphics[height=0.29\textwidth]{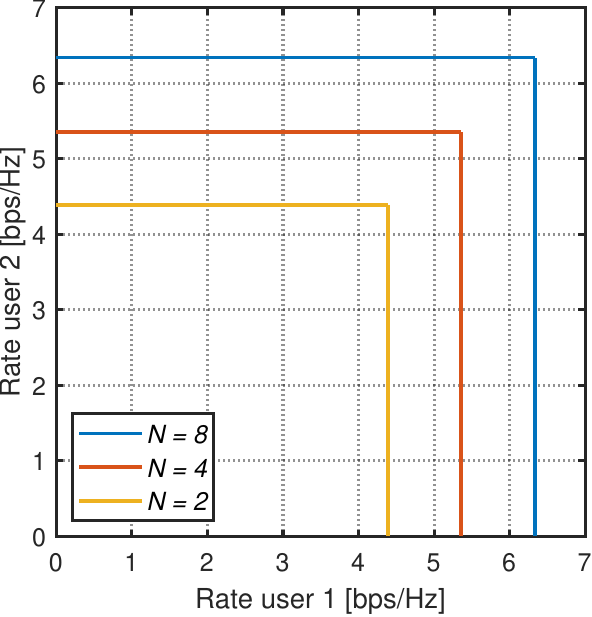}
\caption{Capacity region of a two-user \gls{mac} aided by movable signals, with $P_1/\sigma^2=P_2/\sigma^2=10$~dB.}
\label{fig:mac}
\end{figure}

We have seen that by dynamically setting the signal frequency to $f^\star$ fulfilling \eqref{eq:f} and \eqref{eq:f-neq}, the capacity region is only constrained by $R_k\leq\log_2(1+P_kN/\sigma^2)$, for $k=1,2$.
To exactly achieve $R_k=\log_2(1+P_kN/\sigma^2)$, for $k=1,2$, both users can transmit with their respective powers $P_1$ and $P_2$ on the frequency $f^\star$.
Then, since $f^\star$ makes the two channels $\mathbf{h}_{ul,1}$ and $\mathbf{h}_{ul,2}$ orthogonal, the base station can recover the two transmitted symbols by applying matched filtering to the received signal $\mathbf{y}_{ul}$.
In detail, the base station computes the signal used for detection $\mathbf{z}=[z_1,z_2]^T\in\mathbb{C}^{2\times1}$ as
\begin{equation}
\mathbf{z}=\mathbf{H}_{ul}^H\mathbf{y}_{ul},
\end{equation}
and $z_k$ is used to detect the transmitted signal $x_{ul,k}$ with a rate $R_k=\log_2(1+P_kN/\sigma^2)$.
The resulting capacity region is shown in Fig.~\ref{fig:mac}, where we fix $P_1/\sigma^2=P_2/\sigma^2=10$~dB and consider three different values of $N$.

As a final remark, we observe that the sum rate capacity, defined as the maximum sum rate $R_1+R_2$ achievable in this \gls{mac} system, is given by
\begin{equation}
C_{ul}=
\log_2\left(1+\frac{P_1N}{\sigma^2}\right)+
\log_2\left(1+\frac{P_2N}{\sigma^2}\right).
\end{equation}

\section{Capacity of Broadcast Channel}
\label{sec:bc}

In this section, we derive the capacity region of the two-user \gls{bc} aided by movable signals introduced in Section~\ref{sec:model}.

\subsection{Characterization of the Capacity Region}

An upper bound on the rate of any user is achieved by allocating all the power $P$ to that user as in a single-user \gls{miso} system, namely
\begin{align}
R_k
&\leq\log_2\left(1+\frac{P}{\sigma_k^2}\mathbf{h}_{dl,k}\mathbf{h}_{dl,k}^H\right)\\
&=\log_2\left(1+\frac{P}{\sigma_k^2}N\right),
\end{align}
for $k=1,2$.
By allocating all the power $P$ to user $k$, we obtain $R_k=\log_2(1+PN/\sigma_k^2)$, while the rate of the other user is zero.
These two extreme cases can be bridged by allocating a power $P_1$ to user 1 and $P_2$ to user 2, such that $P_1+P_2=P$, which limits the two rates as
\begin{align}
R_1&\leq\log_2\left(1+\frac{P_1}{\sigma_1^2}N\right),\label{eq:R1}\\
R_2&\leq\log_2\left(1+\frac{P-P_1}{\sigma_2^2}N\right).\label{eq:R2}
\end{align}
Interestingly, this pair of constraints holds tightly when the frequency is reconfigured to orthogonalize the two channels as derived in Section~\ref{sec:mac}, i.e., such that $\mathbf{h}_{dl,1}\mathbf{h}_{dl,2}^H=0$, as shown in the following subsection.
Therefore \eqref{eq:R1} and \eqref{eq:R2} characterize the boundary of the capacity region, where all the points on the boundary can be achieved by using different values of power $P_1$.

\subsection{Achievability of the Capacity Region}

\begin{figure}[t]
\centering
\includegraphics[height=0.29\textwidth]{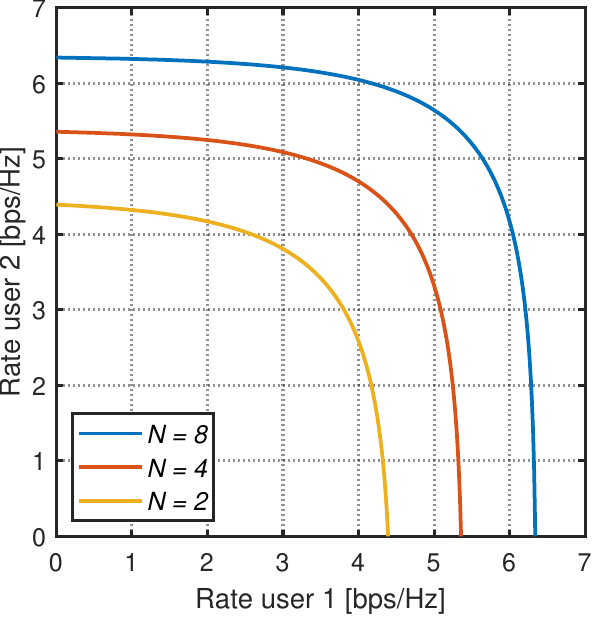}
\caption{Capacity region of a two-user \gls{bc} aided by movable signals, with $P/\sigma_1^2=P/\sigma_2^2=10$~dB.}
\label{fig:bc}
\end{figure}

To achieve the rate pairs $R_1=\log_2(1+P_1N/\sigma_1^2)$ and $R_2=\log_2(1+(P-P_1)N/\sigma_2^2)$, the base station can perform matched beamforming with the appropriate power allocation by operating at a frequency $f^\star$ which fulfills \eqref{eq:f} and \eqref{eq:f-neq}.
In detail, the base station allocates a power $P_1$ to the symbol intended for user 1 $s_1\in\mathbb{C}$, which becomes subject to $\mathbb{E}[\vert s_1\vert^2]=P_1$, and the remaining power to the symbol for user 2 $s_2\in\mathbb{C}$, such that $\mathbb{E}[\vert s_2\vert^2]=P-P_1$.
Then, the base station computes the transmitted signal by precoding the two symbols $\mathbf{s}=[s_1,s_2]\in\mathbb{C}^{2\times1}$ as
\begin{equation}
\mathbf{x}_{dl}=\frac{1}{\sqrt{N}}\mathbf{H}_{dl}^H\mathbf{s},
\end{equation}
where $\mathbf{H}_{dl}=[\mathbf{h}_{dl,1}^T,\mathbf{h}_{dl,2}^T]^T\in\mathbb{C}^{2\times N}$.
Alternatively, we can equivalently interpret the transmitted signal as $\mathbf{x}_{dl}=\mathbf{x}_{dl,1}+\mathbf{x}_{dl,2}$, where $\mathbf{x}_{dl,1}=\mathbf{h}_{dl,1}^Hs_1/\sqrt{N}$ and $\mathbf{x}_{dl,2}=\mathbf{h}_{dl,2}^Hs_2/\sqrt{N}$.
Since $f^\star$ makes the two channels $\mathbf{h}_{dl,1}$ and $\mathbf{h}_{dl,2}$ orthogonal, each user can recover its symbol with no interference from the other stream.
Their rates will therefore be $R_1=\log_2(1+P_1N/\sigma_1^2)$ and $R_2=\log_2(1+(P-P_1)N/\sigma_2^2)$.
The resulting capacity region is shown in Fig.~\ref{fig:bc}, where we consider $P/\sigma_1^2=P/\sigma_2^2=10$~dB and three different values of $N$.

The sum rate capacity is achieved by finding the power allocation $P_1$ for user $1$ that solves
\begin{align}
\underset{P_1}{\mathsf{\mathrm{max}}}\;\;
&\log_2\left(1+\frac{P_1}{\sigma_1^2}N\right)+\log_2\left(1+\frac{\left(P-P_1\right)}{\sigma_2^2}N\right)\label{eq:prob-obj}\\
\mathsf{\mathrm{s.t.}}\;\;\;
&0\leq P_1\leq P.\label{eq:prob-c}
\end{align}
By assuming $\sigma_1^2=\sigma_2^2=\sigma^2$ for simplicity, \eqref{eq:prob-obj}-\eqref{eq:prob-c} can be rewritten as
\begin{align}
\underset{P_1}{\mathsf{\mathrm{max}}}\;\;
&\log_2\left(1+\frac{P}{\sigma^2}N+\frac{P_1\left(P-P_1\right)}{\sigma^4}N^2\right)\\
\mathsf{\mathrm{s.t.}}\;\;\;
&0\leq P_1\leq P,
\end{align}
giving that $P_1$ needs to maximize $P_1(P-P_1)$, which is a concave function maximized in $P_1=P/2$.
By setting $P_1=P_2=P/2$, the sum rate capacity is therefore given by
\begin{equation}
C_{dl}=2\log_2\left(1+\frac{P}{2\sigma^2}N\right).\label{eq:Cdl}
\end{equation}

\section{Frequency Optimization for Movable Signals with Limited Frequency Range}

We have characterized the capacity region of two-user systems aided by movable signals and showed how to optimally select the signal frequency.
While we have assumed so far that the signal frequency can be arbitrarily set, the frequency is selected within a limited frequency range in practice.
In this section, we show how to select the signal frequency when it is confined within a range $[f_{\text{min}},f_{\text{max}}]$, and assess the performance of movable signals under this practical constraint.

Sections~\ref{sec:mac} and \ref{sec:bc} have shown that the rates of the two users are jointly maximized when the two channels are orthogonal, both in the uplink and in the downlink.
Therefore, we optimize the signal frequency so as to minimize the cosine similarity between the two channels.
For the uplink, this is formulated as
\begin{equation}
\underset{f}{\mathsf{\mathrm{min}}}\;\;
\left\vert\mathbf{h}_{ul,2}^H\mathbf{h}_{ul,1}\right\vert\;\;
\mathsf{\mathrm{s.t.}}\;\;
f\in\left[f_{\text{min}},f_{\text{max}}\right].
\end{equation}
while for the downlink the objective is $\vert\mathbf{h}_{dl,1}\mathbf{h}_{dl,2}^H\vert$.
By expressing $\mathbf{h}_{ul,2}^H\mathbf{h}_{ul,1}$ as a function of $\lambda=c/f$ as discussed in Section~\ref{sec:mac}, this problem can be equivalently reformulated as
\begin{equation}
\underset{f}{\mathsf{\mathrm{min}}}\;\;
\frac
{\left\vert 1-e^{j2N f\beta}\right\vert}
{\left\vert 1-e^{j2f\beta}\right\vert}\;\;
\mathsf{\mathrm{s.t.}}\;\;
f\in\left[f_{\text{min}},f_{\text{max}}\right],
\end{equation}
where we introduced $\beta=\pi d_A(\sin(\theta_1)-\sin(\theta_2))/c$, which in turn is equivalent to
\begin{equation}
\underset{f}{\mathsf{\mathrm{min}}}\;\;
\frac
{\left\vert\sin\left(N f\beta\right)\right\vert}
{\left\vert\sin\left(f\beta\right)\right\vert}\;\;
\mathsf{\mathrm{s.t.}}\;\;
f\in\left[f_{\text{min}},f_{\text{max}}\right],
\end{equation}
following the Euler identity.
By analyzing the objective function $\vert\sin(N f\beta)/\sin(f\beta)\vert$, we observe that all its local minima occur at points where the numerator vanishes while the denominator does not.
Consequently, every local minimum corresponds to a zero of the function and is therefore a global minimum.
If at least one solution $f^\star$ satisfying \eqref{eq:f} and \eqref{eq:f-neq} lies within the interval $[f_{\text{min}},f_{\text{max}}]$, that value is selected, since it perfectly orthogonalizes the channels and globally minimizes the objective function.
Otherwise, since no local minimum of $\vert\sin(N f\beta)/\sin(f\beta)\vert$ lies within $[f_{\text{min}},f_{\text{max}}]$, the extreme value theorem implies that the minimum over the interval occurs either in $f_{\text{min}}$ or $f_{\text{max}}$.
We therefore compare $\vert\sin(N f_{\text{min}}\beta)/\sin(f_{\text{min}}\beta)\vert$ and $\vert\sin(N f_{\text{max}}\beta)/\sin(f_{\text{max}}\beta)\vert$, and select $f_{\text{min}}$ if the former is no larger than the latter, while $f_{\text{max}}$ otherwise.

\begin{figure}[t]
\centering
\includegraphics[height=0.30\textwidth]{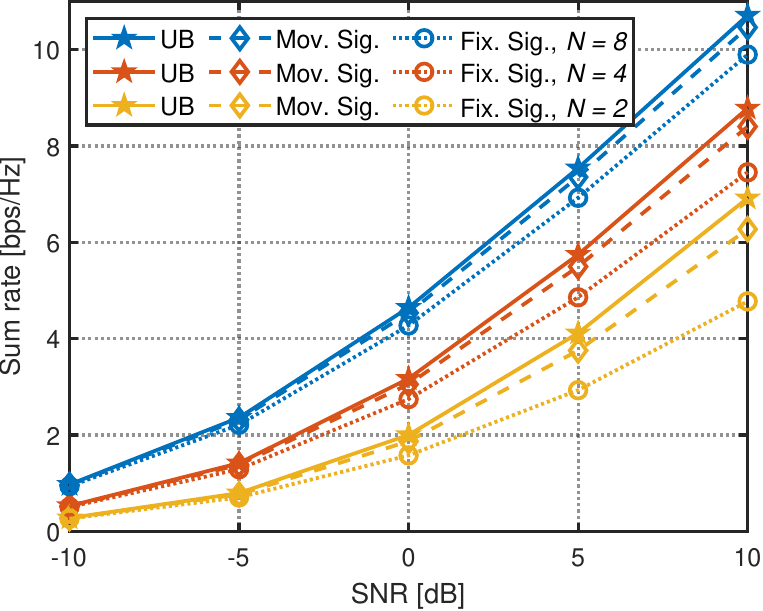}
\caption{Sum rate versus the SNR $P/\sigma^2$.}
\label{fig:sum-rate}
\end{figure}

We now assess the sum rate achieved by movable signals and compare it with its upper bound and the sum rate achievable with fixed signals.
To this end, we consider a \gls{bc} where the users are at distance $d_1=d_2=10$~m from the base station and their angles are independent and uniformly distributed in $[-\pi/2,\pi/2]$, and $f_A=10$~GHz.
In Fig.~\ref{fig:sum-rate}, we report the average sum rate versus the \gls{snr} $P/\sigma^2$ comparing three baselines:
\begin{itemize}
\item ``UB'' denotes the upper bound on the sum rate, given by the sum rate capacity $C_{dl}=2\log_2(1+PN/(2\sigma^2))$, which can be achieved by movable signals with an unconstrained frequency;
\item ``Mov. Sig.'' denotes movable signals with frequency constrained in the range $[f_{\text{min}},f_{\text{max}}]$, with $f_{\text{min}}=f_A$ and $f_{\text{max}}=1.8f_A$, optimized as proposed in this section.
As a precoding strategy, we consider regularized zero-forcing beamforming.
\item ``Fix. Sig.'' denotes fixed signals whose frequency is fixed to $f=f_A$, with regularized zero-forcing beamforming.
\end{itemize}

We make five observations.
\textit{First}, the sum rate always increases with the \gls{snr} and the number of transmitting antennas $N$, as expected from \gls{mimo} theory.
\textit{Second}, movable signals with constrained frequency significantly outperform fixed signals, and nearly achieve the sum rate capacity.
For example, movable signals with constrained frequency improve the sum rate by 31\% over fixed signals, and the sum rate capacity is 45\% higher than the sum rate of fixed signals, when $P/\sigma^2=10$~dB and $N=2$.
The sum rate capacity is not exactly achieved because of the limited frequency range.
\textit{Third}, the gain of movable signals decreases with the number of antennas $N$ since the user channels orthogonalize naturally if the number of antennas is much larger than the number of users.
\textit{Fourth}, the gain of movable signals increases with the \gls{snr}, since their interference-suppression capabilities are more effective in interference-limited scenarios.
\textit{Fifth}, the computational overhead due to movable signals is negligible since the frequency is reconfigured in closed-form.

\section{Conclusion}

We have investigated the fundamental limits of two-user wireless systems aided by movable signals, a recently proposed approach for enabling \gls{sre}.
By dynamically reconfiguring the operating frequency, movable signals can effectively orthogonalize the user channels and suppress inter-user interference.
Therefore, they enable significantly enhanced capacity regions in both \gls{mac} and \gls{bc}.
Future work may extend these results to more complex wireless scenarios, such as with more than two users each having multiple antennas, in \gls{nlos}, with different base station array geometries, and with hardware impairments.

\bibliographystyle{IEEEtran}
\bibliography{IEEEabrv,main}

\begin{thebibliography}{10}
\providecommand{\url}[1]{#1}
\csname url@samestyle\endcsname
\providecommand{\newblock}{\relax}
\providecommand{\bibinfo}[2]{#2}
\providecommand{\BIBentrySTDinterwordspacing}{\spaceskip=0pt\relax}
\providecommand{\BIBentryALTinterwordstretchfactor}{4}
\providecommand{\BIBentryALTinterwordspacing}{\spaceskip=\fontdimen2\font plus
\BIBentryALTinterwordstretchfactor\fontdimen3\font minus \fontdimen4\font\relax}
\providecommand{\BIBforeignlanguage}[2]{{%
\expandafter\ifx\csname l@#1\endcsname\relax
\typeout{** WARNING: IEEEtran.bst: No hyphenation pattern has been}%
\typeout{** loaded for the language `#1'. Using the pattern for}%
\typeout{** the default language instead.}%
\else
\language=\csname l@#1\endcsname
\fi
#2}}
\providecommand{\BIBdecl}{\relax}
\BIBdecl

\bibitem{dir19}
M.~Di~Renzo, M.~Debbah, D.-T. Phan-Huy, A.~Zappone, M.-S. Alouini, C.~Yuen, V.~Sciancalepore, G.~C. Alexandropoulos, J.~Hoydis, H.~Gacanin, J.~de~Rosny, A.~Bounceur, G.~Lerosey, and M.~Fink, ``Smart radio environments empowered by reconfigurable {AI} meta-surfaces: An idea whose time has come,'' \emph{EURASIP Journal on Wireless Communications and Networking}, vol. 2019, no.~1, pp. 1--20, 2019.

\bibitem{dir20}
M.~Di~Renzo, A.~Zappone, M.~Debbah, M.-S. Alouini, C.~Yuen, J.~de~Rosny, and S.~Tretyakov, ``Smart radio environments empowered by reconfigurable intelligent surfaces: How it works, state of research, and the road ahead,'' \emph{IEEE J. Sel. Areas Commun.}, vol.~38, no.~11, pp. 2450--2525, 2020.

\bibitem{wu21}
Q.~Wu, S.~Zhang, B.~Zheng, C.~You, and R.~Zhang, ``Intelligent reflecting surface-aided wireless communications: A tutorial,'' \emph{IEEE Trans. Commun.}, vol.~69, no.~5, pp. 3313--3351, 2021.

\bibitem{won21}
K.-K. Wong, A.~Shojaeifard, K.-F. Tong, and Y.~Zhang, ``Fluid antenna systems,'' \emph{IEEE Trans. Wireless Commun.}, vol.~20, no.~3, pp. 1950--1962, 2021.

\bibitem{zhu24}
L.~Zhu, W.~Ma, and R.~Zhang, ``Modeling and performance analysis for movable antenna enabled wireless communications,'' \emph{IEEE Trans. Wireless Commun.}, vol.~23, no.~6, pp. 6234--6250, 2024.

\bibitem{din25}
Z.~Ding, R.~Schober, and H.~Vincent~Poor, ``Flexible-antenna systems: A pinching-antenna perspective,'' \emph{IEEE Trans. Commun.}, 2025.

\bibitem{ner25-1}
M.~Nerini and B.~Clerckx, ``Enabling smart radio environments in the frequency domain with movable signals,'' \emph{IEEE Trans. Wireless Commun.}, vol.~25, pp. 16\,417--16\,431, 2026.

\bibitem{ner25-2}
M.~Nerini and B.~Clerckx, ``Movable signals with dual-polarized fixed intelligent surfaces: Beyond diagonal reflection matrices,'' \emph{IEEE Commun. Lett.}, vol.~30, pp. 1096--1100, 2026.

\bibitem{she12}
Z.~Shen, A.~Papasakellariou, J.~Montojo, D.~Gerstenberger, and F.~Xu, ``Overview of {3GPP} {LTE}-advanced carrier aggregation for {4G} wireless communications,'' \emph{IEEE Commun. Mag.}, vol.~50, no.~2, pp. 122--130, 2012.

\bibitem{gab23}
Y.~Gabay, N.~Shlezinger, T.~Routtenberg, Y.~Ghasempour, G.~C. Alexandropoulos, and Y.~C. Eldar, ``Wideband {THz} multi-user downlink communications with leaky wave antennas,'' \emph{arXiv preprint arXiv:2312.08833}, 2023.

\bibitem{zhu25}
Z.~Zhu, L.~Wang, X.~Wang, D.~Wang, and K.-K. Wong, ``Spatial-spectral cell-free sub-terahertz networks: A large-scale case study,'' \emph{IEEE Trans. Wireless Commun.}, vol.~24, no.~4, pp. 2956--2967, 2025.

\bibitem{zha21}
B.~Zhai, A.~Tang, C.~Peng, and X.~Wang, ``{SS-OFDMA}: Spatial-spread orthogonal frequency division multiple access for terahertz networks,'' \emph{IEEE J. Sel. Areas Commun.}, vol.~39, no.~6, pp. 1678--1692, 2021.

\bibitem{li22}
R.~Li, H.~Yan, and D.~Cabric, ``Rainbow-link: Beam-alignment-free and grant-free {mmW} multiple access using true-time-delay array,'' \emph{IEEE J. Sel. Areas Commun.}, vol.~40, no.~5, pp. 1692--1705, 2022.

\bibitem{cle13}
B.~Clerckx and C.~Oestges, \emph{MIMO wireless networks: Channels, techniques and standards for multi-antenna, multi-user and multi-cell systems}.\hskip 1em plus 0.5em minus 0.4em\relax Academic Press, 2013.

\end{thebibliography}

\end{document}